\documentclass[twocolumn,showpacs,showkeys,preprintnumbers,amsmath,aps,amssymb,superscriptaddress]{revtex4}
\usepackage[english]{babel}
\usepackage{graphicx}
\usepackage{dcolumn}
\usepackage{bm}
\usepackage{hyperref}
\usepackage{url}
\usepackage{color}
\usepackage{lineno}
\usepackage{amssymb,amsbsy} 
\usepackage{amsthm}
\usepackage{amsmath}
\usepackage{verbatim}
\usepackage{subfigure}
\usepackage{soul}
\usepackage{ulem}
\usepackage{slashed}
\usepackage{rotating}
\usepackage{textcomp}
\usepackage{booktabs}
\usepackage[all]{xy}
\usepackage{cleveref}
\usepackage{ragged2e}

\begin{document}

\setlength{\parindent}{0pt}

\title{The quantum square-well fluid: a thermodynamic geometric view}

\author{J. L. López-Picón} \email{lopezjl@ugto.mx}
\affiliation{Divisi\'on de Ciencias e Ingenier\'ias Campus Le\'on,
    Universidad de Guanajuato, A.P. E-143, C.P. 37150, Le\'on, Guanajuato, M\'exico.}
    \author{L. F. Escamilla-Herrera}
\email{lf.escamilla@ugto.mx}
\affiliation{Divisi\'on de Ciencias e Ingenier\'ias Campus Le\'on,
    Universidad de Guanajuato, A.P. E-143, C.P. 37150, Le\'on, Guanajuato, M\'exico.}
\author{Alejandro Gil-Villegas} \email{gil@fisica.ugto.mx}
\affiliation{Divisi\'on de Ciencias e Ingenier\'ias Campus Le\'on,
    Universidad de Guanajuato, A.P. E-143, C.P. 37150, Le\'on, Guanajuato, M\'exico.}
    \author{Jos\'e Torres-Arenas} \email{jtorres@fisica.ugto.mx}
\affiliation{Divisi\'on de Ciencias e Ingenier\'ias Campus Le\'on,
    Universidad de Guanajuato, A.P. E-143, C.P. 37150, Le\'on, Guanajuato, M\'exico.}
\affiliation{Departamento de Física Aplicada, Universidade de Vigo, Spain}    

\date{\today}

\begin{abstract}
\noindent We investigate several aspects of the thermodynamic geometry for  a quantum fluid with square-well interactions using a third-order perturbation-theory framework based on the path-integral–necklace analogy. A comparison is made between the thermodynamic and geometric properties of the quantum fluid and its classical counterpart for the interaction ranges $\lambda^*=1.3$, 1.5, and 1.7. In particular, we analyze the scalar-curvature behavior, criticality, and the corresponding Widom lines derived from curvature and several thermodynamic response functions. Quantum effects are shown to smooth supercritical anomalies of the scalar curvature and to shift its extrema for short-range interactions, while leaving the critical exponents of both the curvature and its heat capacity consistent with mean-field predictions. Widom lines associated with temperature-dependent response functions and with the curvature scalar exhibit pronounced classical–quantum differences for short interaction ranges; in contrast,  those derived from the isothermal compressibility exhibit only minor variations. Overall, these results highlight the sensitivity of geometric information of thermodynamic systems due to quantum effects and the crucial role of the interaction range in shaping supercritical thermodynamic behavior.
\end{abstract}
 
 \keywords{Statistical Mechanics, Square-well fluid, Thermodynamic Geometry}
\pacs{05.20.Jj, 95.30.Sf, 95.30.Tg}

\maketitle


\section{Introduction}

For certain substances such as hydrogen, deuterium, helium and neon, it has been well known for decades that it is necessary to consider quantum contributions to model the behavior of the molecules to accurately predict their thermodynamic and structural properties \cite{Kim1969,nordholm1993statistical, Yoon1988, sadus1992influence,  Sese1995,Ceperley1995,Wang1997,Wolbach1997,Noya2011,Trejos2012,Trejos2013,Muller2019, Gil-Villegas2024, Mejia2025}. For the sake of simplicity, throughout this work we will refer to fluids exhibiting these features as \textit{quantum fluids}.

Quantum fluids represent a distinctive class of systems in which atomic or molecular motion cannot be accurately described within a classical framework. Even when effective  pairwise additive potentials are employed to model electrostatic interactions, nuclear motion must still be treated from a non-classical framework.\\
Quantum effects become significant when the thermal de Broglie wavelength, $\lambda_{_{B}} = h/\sqrt{2m\pi k_B T}$ (where $h$ is the Planck constant, $k_B$ the Boltzmann constant, $m$ the particle mass and $T$ stands for temperature),  is comparable to the characteristic interparticle spacing, thereby the classical description in such cases is inadequate. This regime becomes particularly relevant for substances with low-mass particles, such as those found in hydrogen, deuterium, helium, and neon,  where phenomena such as zero-point motion and quantum tunneling could play a dominant role \cite{Prisk2023,Bell2023}.

Quantum fluids offer practical applications in areas such as energy storage and cryogenic transport. Such systems provide a valuable testing ground for exploring quantum effects in dense, interacting media and contribute to a deeper understanding of matter under extreme thermodynamic conditions \cite{Fang2019,Skrbek2024}. 

The study of quantum fluids has progressed through both theoretical analyzes and 
computational simulations. A variety of frameworks have been formulated to 
characterize these systems, including perturbation theory, Gaussian wave packets, 
quantum Monte Carlo techniques, and the path integral (PI) formalism 
\cite{singh1977quantum, singh1978quantum, Zhao2004,Kim2008,Georgescu2013,nordholm1993statistical, Trejos2012, Trejos2013, Muller2019, serna2016molecular,Fantoni2014,Li2025}.\\
One possible route to obtain quantum predictions in thermodynamic systems involves introducing corrections to classical expressions. Under this approach, many of the seminal works exploring the thermodynamic properties of systems with quantum considerations used semiclassical approximations\cite{wigner1932quantum,kirkwood1933quantum,green1951quantum}.\\
In early efforts to model quantum systems, the hard sphere model has played a very important role, first as an individual quantum fluid system but also in its role as a reference system in perturbation theories\cite{derderian1971expansion,Kim1969,barker1979quantum,gibson1975quantum,gibson1975quantumII,singh1977quantum, serna2016molecular}. On the other hand, the Feynman–Hibbs path-integral method represents each particle as a polymer-like ring composed of classical beads connected by harmonic springs. The corresponding path integrals can be evaluated using Monte Carlo, molecular dynamics, or Brownian dynamics techniques. This technique allows to obtain precise quantum description of the studied systems.\\ 
In recent years, the Feynman–Hibbs path-integral method, thermodynamic perturbation theory and the SAFT-VR approach \cite{Gil-Villegas1997} have been successfully applied to the quantum Square-Well (SW) fluid \cite{serna2016molecular}, one of the simplest yet most informative models in Statistical Mechanics. Defined  by a hard-core repulsion and a finite-range attractive well, the SW potential captures essential features of intermolecular interactions while remaining analytically and computationally tractable. Its piecewise definition allows for a clear separation between repulsive and attractive contributions, making it an ideal framework for assessing theoretical developments and exploring phase behavior. Consequently, the SW fluid has been extensively used to study liquid–gas coexistence, critical phenomena, and the thermodynamic properties of simple and associating fluids \cite{Benavides1989,Chang1994, GilVillegas1996,Scholl2005, Gil-Villegas1997,DelRio2002, Patel2005}. In the present study, we employ a Helmholtz free energy expression formulated for the SW fluid, derived within the framework of perturbation theory to explore the thermodynamic geometry of a quantum square-well fluid. The analysis extends the perturbation expansion up to third-order, with the quantum hard-sphere system serving as the reference system and quantum corrections explicitly included in the higher-order contributions.

The quantum fluid under consideration consists of $N$ hard spheres of diameter $\sigma$ and mass $m$ that interact via a square-well potential of depth $\epsilon$ and range $\lambda$,  
\begin{equation} \label{eq:1}
	\phi (r) =
	\begin{cases}
	\infty \,, & r< \sigma \,, \\
	-\epsilon \,, & \sigma \le r \le \lambda \sigma \,, \\
	0 \,, & r > \lambda\sigma \,.  \\
	 \end{cases}
\end{equation}

The aforementioned equation of state (EOS) was computed using thermodynamic perturbation theory up to the third-order, according to a Zwanzig expansion in $\beta = 1/k_{B}T$ \cite{zwanzig1954high},
\begin{equation}\label{QAEOS}
\frac{A}{Nk_{_{B}}T} = \frac{A^{\text{ideal}}}{Nk_{_{B}}T} + \frac{A^{\text{QHS}}}{Nk_{_{B}}T} + a_{1_{\text{PI}}}\beta + a_{2_{\text{PI}}}\beta^2 + a_{3_{\text{PI}}}\beta^3\,.   
\end{equation}
where $N$ is the number of particles, $A$ and $A^{\text{QHS}}$ are the Helmholtz free energies of the quantum fluid and the Quantum Hard Sphere (QHS) reference system, respectively; $A^{\text{ideal}}$ is the ideal contribution, and $a_{nPI}$, the perturbation terms. The QHS structural properties required for the evaluation of the perturbation terms were obtained from MC simulations using the exact analogy between the discretized path-integral formulation of quantum mechanics and the partition function of a fluid composed of ring molecules \cite{chandler1981exploiting}.
The SW parameters $\epsilon$ and $\sigma$ are used to define the reduced variables of temperature, $T^{*} = k_{_B}T/ \epsilon$, pressure, $P^{*} = P \sigma^3/\epsilon$, packing fraction $\eta = \pi \sigma^3 \rho/6$, potential range, $\lambda^*=\sigma\lambda$, and the de Boer parameter $\Lambda$, 
\begin{equation}
\Lambda = \frac{h}{\sigma \sqrt{m\epsilon}}\,;
\end{equation}
In this work, we will set this parameter to $\Lambda = \sqrt{2\pi}/4$, which is close to the value of neon (Ne). We  investigate the influence of quantum effects on the thermodynamic geometry of the square-well fluid. This is achieved through a comparative analysis with the corresponding classical system.

This work is organized as follows. In Section II,  theoretical expressions of thermodynamic geometry are briefly reviewed, and present the classical and quantum equations of state for the SW system considered in this work. Section III presents the results of our analysis for both the subcritical and supercritical regimes, emphasizing the key differences identified between them highlighting the construction of Widom lines; additionally, critical exponents for all response functions and curvature scalar $R$ are calculated,  presented and discussed. Section IV concludes with a summary of the main findings of this work.


\section{Theoretical Methods}
\subsection{Thermodynamic Geometry}

This work is based on the theoretical framework of thermodynamic geometry (TG) developed by G. Ruppeiner \cite{Ruppeiner1979,Ruppeiner1995}, which is derived from  entropic representation  and Einstein-Landau thermodynamic fluctuation theory \cite{Einstein1910, LandauLifshitz1980}. This  framework is similar to the one previously constructed by Weinhold in energy representation \cite{Weinhold1975}. Although  the derived TG metric on  the thermodynamic equilibrium space is naturally a metric in entropic representation, it is much more common and convenient  to transform this metric into the Helmholtz free energy $A$ representation due to its particular simplicity, because for this potential the metric $[g_{ij}]_A$ becomes diagonal. In terms of energy divided by volume $f = A/V$, the metric is simply,
\begin{equation} \label{TG:metric}
[g_{ij}]_A = \frac{1}{k_B T}\ \left(
\begin{array}{cc}
-  \frac{\partial^2 f}{\partial T^2}     &  0\\
0     & \frac{\partial^2 f}{\partial \rho^2}
\end{array} \right)\,;
\end{equation}
or, equivalently, in terms of response functions,
\begin{equation*}
[g_{ij}]_A = \frac{1}{k_B}\ \left(
\begin{array}{cc}
\frac{c_V}{T^2}     &  0\\
0     & \frac{1}{T\rho^2\kappa_T}
\end{array} \right)\,,
\end{equation*}
where $c_V$ is the constant‑volume heat capacity per unit volume, $\rho = N/V$ is the number density, and $\kappa_T$ is the isothermal compressibility.\\
For systems written in terms of other thermodynamic potentials with two thermodynamic coordinates $\left\{X,Y\right\}$ (a two-dimensional manifold), the scalar curvature $R$ associated with an arbitrary metric $g_{ij}$  not necessarily diagonal,  has the general form:
\begin{align}
R = & -\frac{1}{\sqrt{g}} \frac{\partial}{\partial X}\left(  \frac{g_{_{XY}}}{g_{_{XX}}\sqrt{g}} \frac{\partial g_{_{XX}}}{\partial Y} - \frac{1}{\sqrt{g}} \frac{\partial g_{_{YY}}}{\partial X} \right)  \\ \nonumber
& - \frac{1}{\sqrt{g}} \frac{\partial}{\partial Y}\left(  \frac{2}{\sqrt{g}} \frac{\partial g_{_{XY}}}{\partial X} - \frac{1}{\sqrt{g}} \frac{\partial g_{_{XX}}}{\partial Y} - \frac{g_{_{XY}}}{g_{_{XX}}\sqrt{g}} \frac{\partial g_{_{XX}}}{\partial X} \right)\,, 
\end{align}
which simplifies considerably if the metric is diagonal as in eq.\ref{TG:metric}.

Regarding the explicit EOS considered in this work, the function $f$ can be written in terms of the thermodynamic function $a = \frac{A}{Nk_{_{B}}T}$,  
\begin{equation}
f = \rho k_{B}Ta\,.
\end{equation}
For simplicity, in the following, all  thermodynamic and geometric quantities are expressed in terms of the reduced set of variables $T^{*}$ and  $\rho^{*}$. For this particular set, the components of the metric are:
\begin{align}  \label{TG:components}
g_{\text{\tiny TT}} &= \frac{1}{ \sigma^3}\left( - \rho^{*} \frac{\partial^2 a}{\partial T^{*2}} - \frac{2 \rho^{*}}{T^{*}} \frac{\partial a}{\partial T^{*}}  \right)  = \frac{1}{\sigma^{3}} G_{_{TT}}(\rho^{*},T^{*})\,, \nonumber \\
g_{\rho\rho} &= \sigma^3 \left( \rho^{*} \frac{\partial^2 a}{\partial \rho^{*2}} + 2 \frac{\partial a}{\partial \rho^{*}}  \right) = \sigma^{3}G_{\rho \rho}(\rho^{*},T^{*})\,.
\end{align}

The scalar curvature has dimensions of cubed distance, and  its reduced expression $R^{*} = R/\sigma^3$ is given by,  
\begin{align} \label{TG:redscalar}
R^{*} = &  \frac{1}{\sqrt{g}} \frac{\partial}{\partial T^{*}}\left( \frac{1}{\sqrt{g}} \frac{\partial G_{\rho \rho}}{\partial T^{*}} \right) + \frac{1}{\sqrt{g}} \frac{\partial}{\partial \rho^{*}} \left(  \frac{1}{\sqrt{g}} \frac{\partial G_{_{TT}}}{\partial \rho^{*}} \right)\,.
\end{align}

Finally, to be able to have a straightforward comparison between the thermodynamic quantities reported in Ref. \cite{serna2016molecular}, we use the packing fraction $\eta$ instead of the number density in our calculations.


\subsection{Equations of State}

In this work, we consider the Helmholtz free energy for both a  classical square-well fluid and its quantum counterpart, which incorporates quantum-mechanical contributions. In both cases, the free energy expressions are  obtained via perturbation theory. An important feature of  quantum contributions is their explicit dependence on temperature through the de Broglie thermal wavelength $\lambda_{{_B}}$.

\subsubsection{Classical SW Helmholtz free energy}

For the classical SW fluid, the reference properties are obtained from classical Hard Spheres (HS), and the HS free energy is obtained from the Carnahan-Starling (CS) EOS \cite{Carnahan69},
\begin{equation}\label{CSEOS}
\frac{A^{\text{HS}}}{Nk_{_{B}}T} = \frac{4\eta -3\eta^2}{(1 -\eta)^2}\,.
\end{equation}
Only first- and second-order perturbative terms are considered, according to the EOS by Patel and coworkers \cite{Patel2005}. The first‑order correction is given by, 
\begin{equation}
\frac{A_1}{Nk_{_{B}}T} = -4\left( \frac{\epsilon}{k_{_{B}}T}  \right) (\lambda^{*3}-1)\eta g^{\text{HS}}(1;\eta_{\text{eff}})\, ,
\end{equation}
where $g^{\text{HS}}$ is the  HS contact radial distribution function evaluated at an effective packing fraction $\eta_{eff}$,,
\begin{equation}
g^{\text{HS}}(1;\eta_{\text{eff}}) = \frac{1-\eta_{\text{eff}}/2}{(1-\eta_{\text{eff}})^3}; \quad \eta_{\text{eff}} = \frac{c_1\eta + c_2\eta^2}{(1+c_3\eta)^3}\,;
\end{equation}
where $c_1,c_2$ and $c_3$ are given by the following polynomials,
\begin{align}
& c_1 = -\frac{3.1649}{\lambda^{*}} + \frac{13.3501}{\lambda^{*2}} - \frac{14.8057}{\lambda^{*3}}  + \frac{5.7029}{\lambda^{*4}}\,, \\ \nonumber
& c_2 = \frac{43.0042}{\lambda^{*}} - \frac{191.6623}{\lambda^{*2}} - \frac{273.8968}{\lambda^{*3}}  - \frac{128.9334}{\lambda^{*4}}\,, \\ \nonumber
& c_3 = \frac{65.0419}{\lambda^{*}} - \frac{266.4627}{\lambda^{*2}} - \frac{361.0431}{\lambda^{*3}}  - \frac{162.6996}{\lambda^{*4}}\,.
\end{align}

The second-order perturbation term is expressed as
\begin{equation}
\frac{A_2}{Nk_{_{B}}T} = \frac{1}{2}\left( \frac{\epsilon}{k_{_{B}}T}  \right) K^{\text{HS}} \eta \frac{\partial}{\partial \eta} \left( \frac{A_1}{Nk_{_{b}}T} \right)\,,
\end{equation}
where the CS HS isothermal compressibility $K^{\text{HS}}$ is given by \cite{Carnahan69}
\begin{equation}
K^{\text{HS}} = \frac{(1-\eta)^4}{1 + 4\eta + 4\eta^2 + 4\eta^3 \eta^4}\,.
\end{equation}

\subsubsection{Quantum SW (QSW) Helmholtz free energy}

The QSW Helmholtz free energy is given by Eq.(\ref{QAEOS}), using the quantum equation of state derived by Serna and Gil-Villegas \cite{serna2016molecular}. The expression for $A^{QHS}$ is the same as the CS equation of state, Eq.(\ref{CSEOS}), evaluated at the packing fraction 
$$\eta_{QHS} =(1+d_1\lambda_{_B}^*)\eta + (d_2\lambda_{_B}^* + d3\lambda_{_B}^{*2})\eta^2,
$$
with $d_1 = 1.6593854484$, $d_2=-1.0927115150$ and $d_3=-1.1188233921$.

The corresponding first perturbation  term is given by,
\begin{equation}
a_{1_\text{PI}} = -4\eta (\lambda_{\text{eff}}^3-1)g_{\text{HS}}\,,
\end{equation}
where $\lambda_{\text{eff}}$ is an effective range parameter and
\begin{align}
& g_{\text{HS}} = \frac{1-\eta_{\text{eff}}/2}{(1-\eta_{\text{eff}})^3}\,, \\
& \eta_{\text{eff}} = c_1(\lambda^*)\eta + c_2(\lambda^*)\eta^2 + c_3\eta^3\,.
\end{align}
The effective range $\lambda_{\text{eff}}$ is given by the following expressions:
\begin{equation}
\lambda_{\text{eff}} = q_0 + q_1\eta + q_2 \eta^2\,;
\end{equation}
where, 
\begin{align}
& q_0(\lambda^*) = q_{00} + q_{01}\lambda^*\,, \\
& q_1(\lambda^*) = q_{10} + q_{11}\lambda^* + q_{12}\lambda^{*2} + q_{13}\lambda^{*3}\,, \\
& q_2(\lambda^*) = q_{20} + q_{21}\lambda^* + q_{22}\lambda^{*2} + q_{23}\lambda^{*3}\,.
\end{align}
and,
\begin{equation} \label{lambdaeff}
q_{nm}(\lambda_{_{B}}) = q_{nm0} + q_{nm1}\lambda_{_{B}} + q_{nm2}\lambda^2_{_{B}} + q_{nm3}\lambda^3_{_{B}}\,,
\end{equation}
The coefficients $q_{ij}$ are given in Table I.

\begin{table}[ht]
\centering
\begin{tabular}{|c||c|c|c|c|}
 \hline
$nm$ & $q_{nm0}$ & $q_{nm1}$ & $q_{nm2}$ & $q_{nm3}$ \\ \hline \hline
$00$ & $0.0$  & -0.5207594172 & 0.1376491457 & 0.1619400037 \\ \hline
$01$ & 1.0 & 0.2085532937 & -0.09789365585 & -0.09331727334   \\ \hline
$10$ & 0.0 & 27.87555242 &  -16.84467685 & 24.12334316 \\ \hline
$11$ & 0.0 & -61.37020515 & 35.15220145 & -49.99085495  \\ \hline
$12$ & 0.0 & 43.76848829 &  -23.81353044 & 33.33648824  \\ \hline
$13$ & 0.0 & -9.887151471 &  5.374121467 & -7.222508701   \\ \hline
$20$ & 0.0 & -175.6647412 &  -0.4555948621 & 14.16499116  \\ \hline
$21$ & 0.0 & 379.0986356 & -60.59299424 & 2.185720582   \\ \hline
$22$ & 0.0 & -264.2859871 & 83.30714552 & -23.29735407  \\ \hline
$23$ & 0.0 & 59.32562146 & -27.68449440 & 9.881196243  \\ \hline
\end{tabular}
 \caption{Parameters of the effective attractive range $\lambda_{\text{eff}}$ used to describe the quantum mean attractive energy in Eq. (\ref{lambdaeff}).}
\vspace{0.1cm}
\end{table}

The second-order term is given by  \cite{gil1993properties}
\begin{equation}
a_{2_\text{PI}}= \frac{1}{2}K_{\text{HS}}Ya_{1_{\text{PI}}}\,;
\end{equation}
where,
\begin{align}
K_{\text{HS}} = \frac{(1-\eta)^4}{1+4\eta+4\eta^2 -4\eta^3-\eta^4}, \\
Y = \frac{1+\Omega_1 \eta + \Omega_2 \eta^2 + \Omega_3 \eta^3 + \Omega_4 \eta^4}{(1-\eta)^2},
\end{align}
and,
\begin{align} \label{Omegas}
\Omega_{i} = \Omega_{i0} + \Omega_{i1}\lambda_{\text{eff}} + \Omega_{i2}\lambda_{\text{eff}}^2 + \Omega_{i3}\lambda_{\text{eff}}^3\,;
\end{align}
The coefficients $\Omega_{ij}$ ($j=1,2,3$) are given in Table II.

\begin{table}[h]
\centering
\begin{tabular}{|c||c|c|c|c|}
 \hline
$n$ & $\Omega_{n0}$ & $\Omega_{n1}$ & $\Omega_{n2}$ & $\Omega_{n3}$ \\ \hline \hline
$1$ & $-287.2698$  & 564.4165 & -360.5291 & 74.58325 \\ \hline
$2$ & 2374.631 & -4312.958 & 2557.982 & -494.2050 \\ \hline
$3$ & -605.3505 & -105.1644 & 885.7259   & -360.7473 \\ \hline
$4$ & -10317.90 & 21338.96 & -14497.67  & 3243.0211   \\ \hline
\end{tabular}
  \caption{Parameters for the evaluation of the $Y$ function in the  generalized compressibility approximation in Eq. (\ref{Omegas}).}
\vspace{0.1cm}
\end{table}
Finally, the third-order perturbation term is
\begin{equation}
a_{3_\text{PI}} = \frac{1}{6} K^2_{\text{HS}}(1-S\eta)a_{1_{\text{PI}}}\,;
\end{equation}
\noindent where
\begin{equation}
S = \frac{49\lambda^5_{\text{eff}}+ 49\lambda^4_{\text{eff}}- 293\lambda^5_{\text{eff}}-5\lambda^2_{\text{eff}}+211 \lambda_{\text{eff}}+211 }{4(\lambda^2_{\text{eff}}+\lambda_{\text{eff}}+1)}\,.
\end{equation}

Tables III and IV show, respectively, the classical and quantum critical values  of temperature, packing fraction, and pressure for specific SW ranges $\lambda^* = 1.3, 1.5, 1.7$, where it is known that both classical and quantum SW EOS are accurate. The critical values were calculated using standard thermodynamic relations using the equations of state already described. 

\begin{table}[ht]
  \centering
  \begin{tabular}{|c|c|c|c|}
    \hline
    $\lambda^{*}$ & $T^{*}_{\mathrm{cr}}$ & $\eta_{\mathrm{cr}}$ & $P^{*}_{\mathrm{cr}}$ \\ \hline\hline
    1.3 & 0.940934 & 0.206423 & 0.160563 \\ \hline
    1.5 & 1.32907  & 0.149937 & 0.143413 \\ \hline
    1.7 & 1.76812  & 0.125244 & 0.150339 \\ \hline
  \end{tabular}
    \caption{Critical values for different potential ranges of the classical square-well fluid.}
\end{table}

\begin{table}[h]
  \centering
  \begin{tabular}{|c|c|c|c|}
    \hline
    $\lambda^{*}$ & $T^{*}_{\mathrm{cr}}$ & $\eta_{\mathrm{cr}}$ & $P^{*}_{\mathrm{cr}}$ \\ \hline \hline
    1.3 & 0.639694  & 0.157293 & 0.0791678  \\ \hline
    1.5 & 0.983404 &  0.127644 & 0.0796721  \\ \hline
    1.7 & 1.4559 & 0.111839 & 0.109786 \\ \hline
  \end{tabular}
    \caption{Critical values for different potential ranges of the quantum square-well fluid.}
\end{table}

Using these EOS, the thermodynamic geometric curvature and its scalar $R$ are calculated for the classical and quantum square-well equations, in order to understand the role that quantum contributions play in the thermodynamic geometry of the square-well system. An analysis of the behavior of the Widom lines of thermodynamic response functions is also given: thermal expansion coefficient $\alpha$, isothermal compressibility  $\kappa_T$ and heat capacity at constant pressure $C_P$, as well as the geometric $R$-Widom line given by the curvature extrema in the supercritical region.


\section{Results}

To provide a detailed comparison of the thermodynamic geometry exhibited by classical and quantum square-well fluids, we focus on three representative potential ranges, namely $\lambda^* = 1.3, 1.5,$ and $1.7$, which allow us to systematically assess the influence of the interaction range on the resulting geometric behavior.\\
Our EOS  have a restricted domain of validity for both density and temperature. With respect to temperature, their validity is limited, especially in the subcritical region. Therefore, we considered a lower temperature limit based on an estimate of the triple point. The triple point temperature of the SW fluid is known to increase with the potential range\cite{Liu2005}. The  lower limit for temperature  considered in this work for the analyzed ranges is: for $\lambda^* = 1.3$, $T^{*}_{\text{min}} = 0.7$; for $\lambda^* = 1.5$, $T^{*}_{\text{min}} = 0.5$; and for $\lambda^* = 1.7$, $T^{*}_{\text{min}} = 0.4$.

\begin{figure}[h]
\centering
\includegraphics[width=\columnwidth]{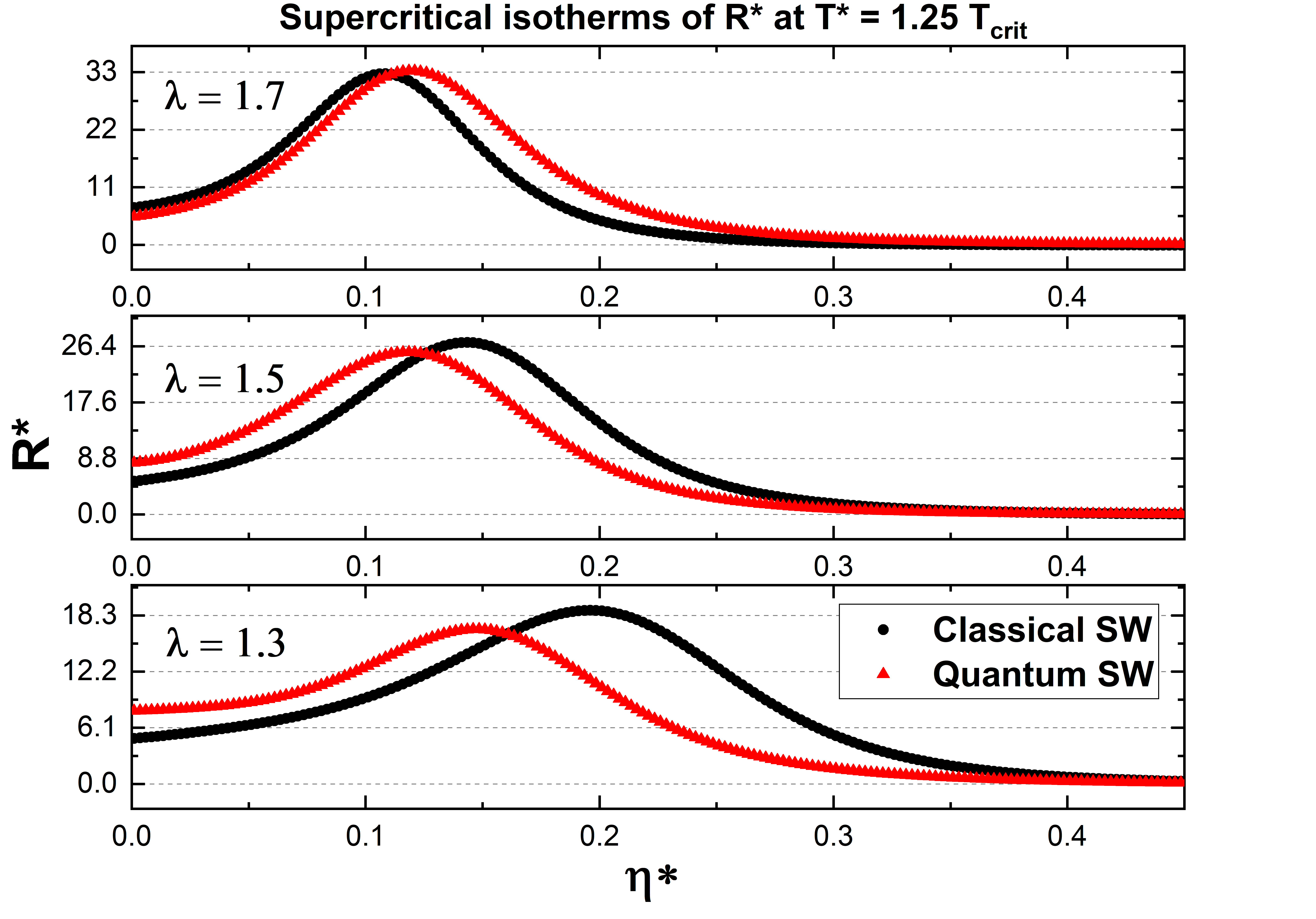}\\
\hrulefill \\
\caption{Comparison of the classical and quantum supercritical isotherms of the reduced scalar curvature  $R^*$ at a temperature $T^* = 1.25  T_{c}$, for $\lambda^* = 1.3$ (bottom), $\lambda^* = 1.5$ (middle) and $\lambda^* = 1.7$ (top).}
\label{Rsuper}
\end{figure}

Figure \ref{Rsuper} presents the classical and quantum isotherms of the reduced scalar curvature $R^*$ at $T/T_c = 1.25$ for the ranges of selected parameters, spanning  supercritical states throughout the full density domain. In this regime, quantum effects smoothen the characteristic anomalies present near the critical point and shift the extrema toward lower densities for short-range potentials.

\begin{figure}[ht]
\centering
\includegraphics[width=\columnwidth]{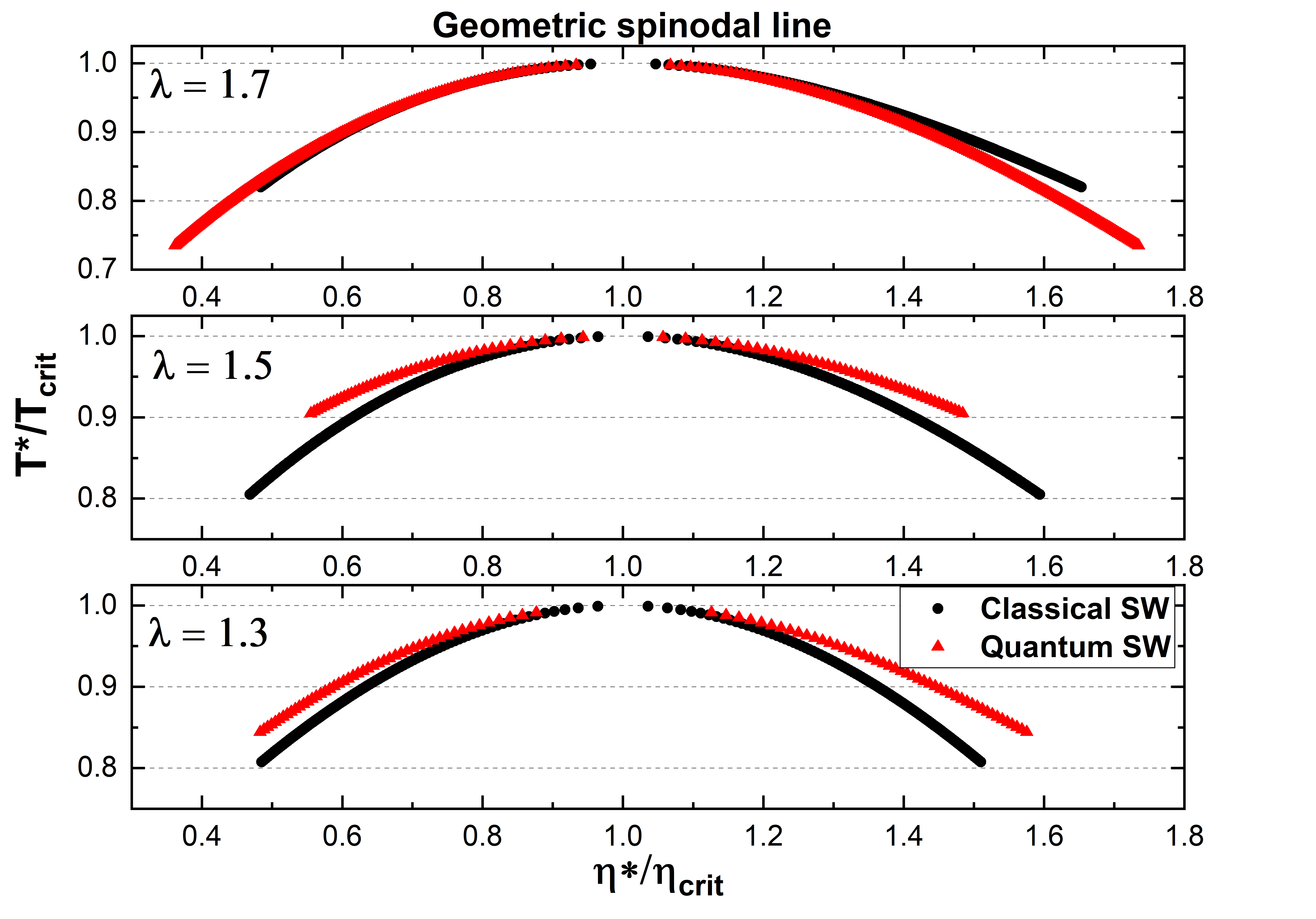}\\
\hrulefill \\
\caption{Comparison of  the Classical and Quantum geometric spinodal lines for $\lambda^* = 1.3$ (bottom), $\lambda^* = 1.5$ (middle), and $\lambda^* = 1.7$ (upper).}
\label{spinodal}
\end{figure}

The region of thermodynamic stability can be determined directly from the curvature scalar $R$.  The spinodal curve is obtained from the divergences exhibited by $R$ in the subcritical region, or equivalently from the zeros of its inverse. Applying this last criteria, the spinodal curves for the interaction ranges studied in this work are presented in Figure~\ref{spinodal}. As can be observed, quantum effects enlarge the stability region relative to the classical case for short interaction ranges of the potential, tending to converge as the interaction range increases.

An initial question addressed here is whether the quantum contributions encoded in the free-energy given in Eq. \eqref{QAEOS} could modify  the behavior of the fluid in the vicinity of the critical point. Although such modifications were not anticipated in light of the universality of critical phenomena for fluids, the approximate and  perturbative nature of the present theory motivated us to examine this issue explicitly.

\begin{figure}[h]
\centering
\includegraphics[width=\columnwidth]{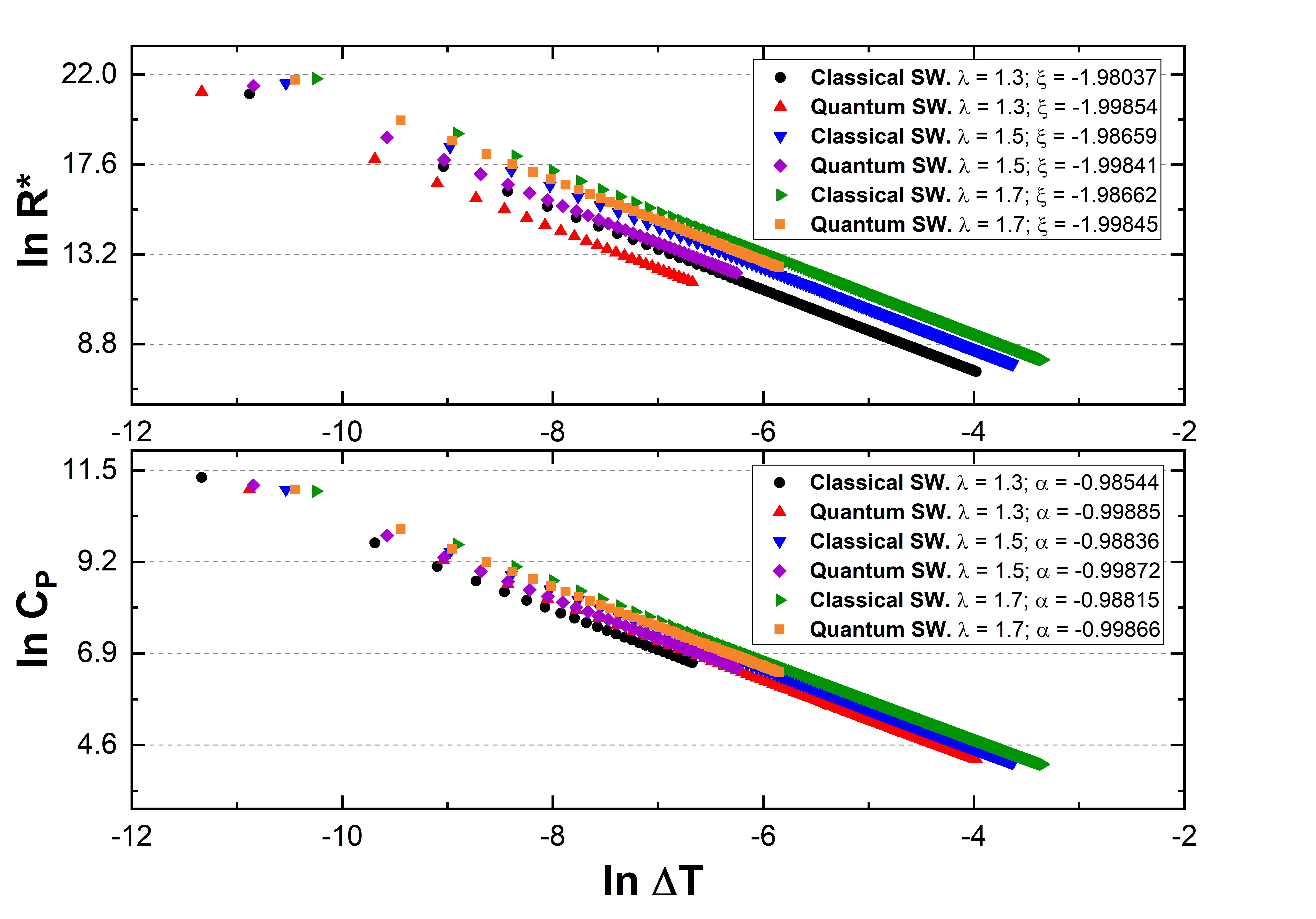}\\
\hrulefill \\
\caption{Critical behavior  for the classical and quantum square-well curvature $R^{*}$ (upper) and $C_P$ (bottom). Both quantities are evaluated  along the critical isochore ($\eta = \eta_{\text{cr}}$), as temperature approaches  the critical point from above  from  $102\%$ to  $100.002\%$  its critical value for  $\lambda^* = 1.3,1.5,1.7$.}
\label{criticality}
\end{figure}

As shown in Figure \ref{criticality},the critical exponents associated to $R$ and $C_p$ for the classical and quantum square-well fluids, are obtained from the expected linear behavior  in a log–log plot of the response function versus $\Delta T = T - T_c$ to simplify the numerical approach, reducing computational errors in the iterative process performed. 
This behavior is obtained via the following relation,
\begin{equation}
    \ln\left[f_{\text{res}}(T)\right]= \beta \ln|\Delta T| + a\,;
\end{equation}
where $f_{\text{res}}(T)$ stands for the thermodynamic function considered, $\beta$ is its corresponding critical exponent, and $a$ is the $y$-intercept, which is an irrelevant offset to the critical behavior of $f_{\text{res}}(T)$. This equation is applied approaching from 102\% the value of the critical temperature up to 100.002\% for $\eta = \eta_{\text{cr}}$; we approach the critical point from the supercritical region along the critical isochore.

The critical exponents calculated  for $R^{*}$ and $C_P$,  equal to $2$ and $1$ respectively, are fully consistent with mean-field predictions, in line  with the perturbative nature of the theory. Consistently, the results are identical for both the classical and quantum versions of the theory. The behavior near the critical point therefore follows a mean-field description that neglects fluctuations in the thermodynamic variables.

\begin{figure}[h]
\centering
\includegraphics[width=\columnwidth]{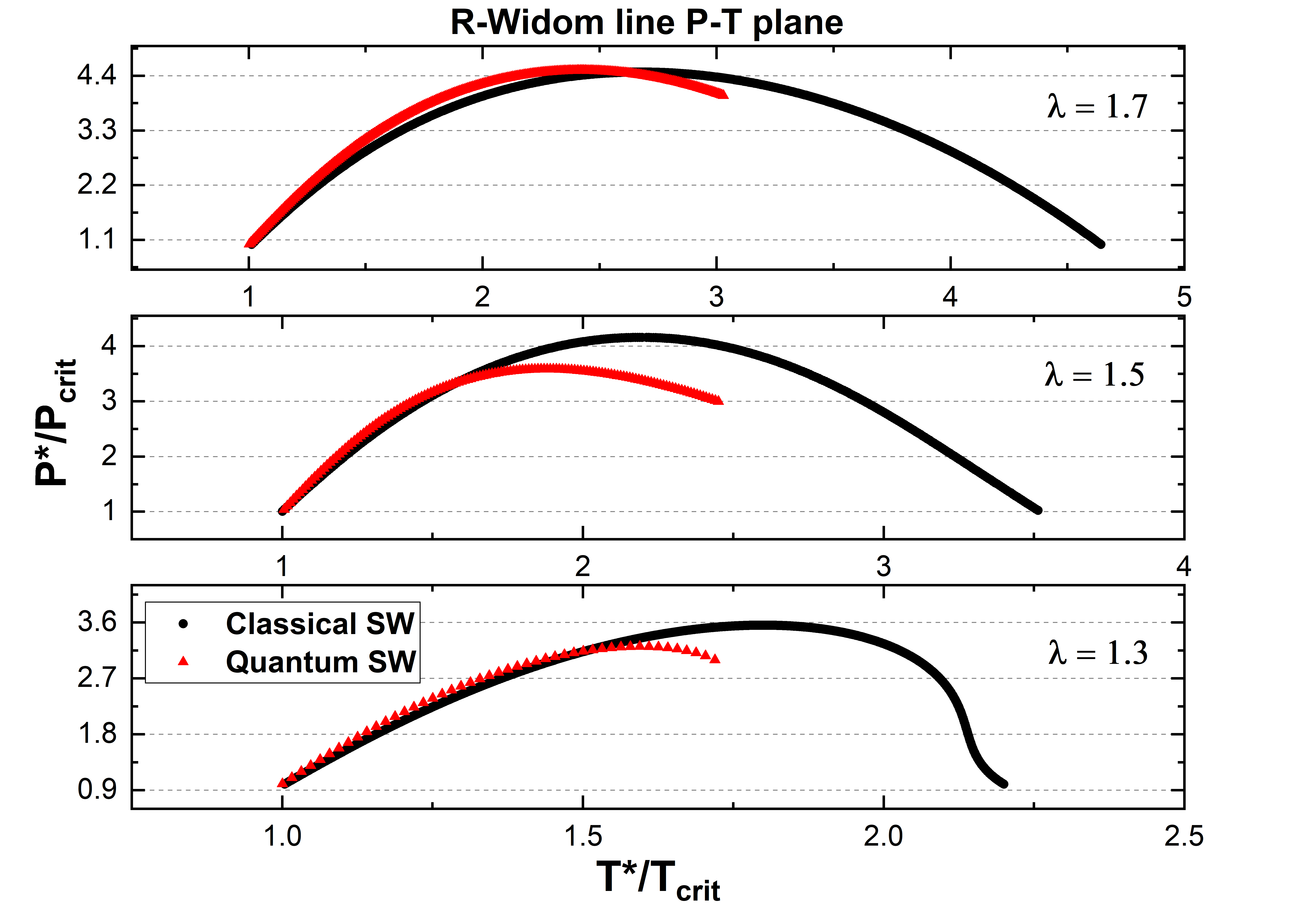}\\
\hrulefill \\
\caption{Comparison of  the classical and quantum $R$-Widom lines for  $\lambda^* = 1.3$ (bottom), $\lambda^* = 1.5$ (middle), and $\lambda^* = 1.7$ (top).}
\label{RWidom}
\end{figure}

The classical and quantum $R$-Widom lines are presented in Figure \ref{RWidom}. As shown, noticeable differences between the classical and quantum behaviors persist across all ranges considered. Because the scalar curvature involves higher-order derivatives with respect to temperature and density, it is reasonable to attribute  these discrepancies to this intrinsic feature of the thermodynamic quantity. For the ranges examined, the convergence between the classical and quantum results observed for other Widom lines (see below) as the range increases is not apparent in this case. This suggests that any such convergence, if present, requires larger ranges outside of the validity domain of the considered EOS.

\begin{figure}[h]
\centering
\includegraphics[width=\columnwidth]{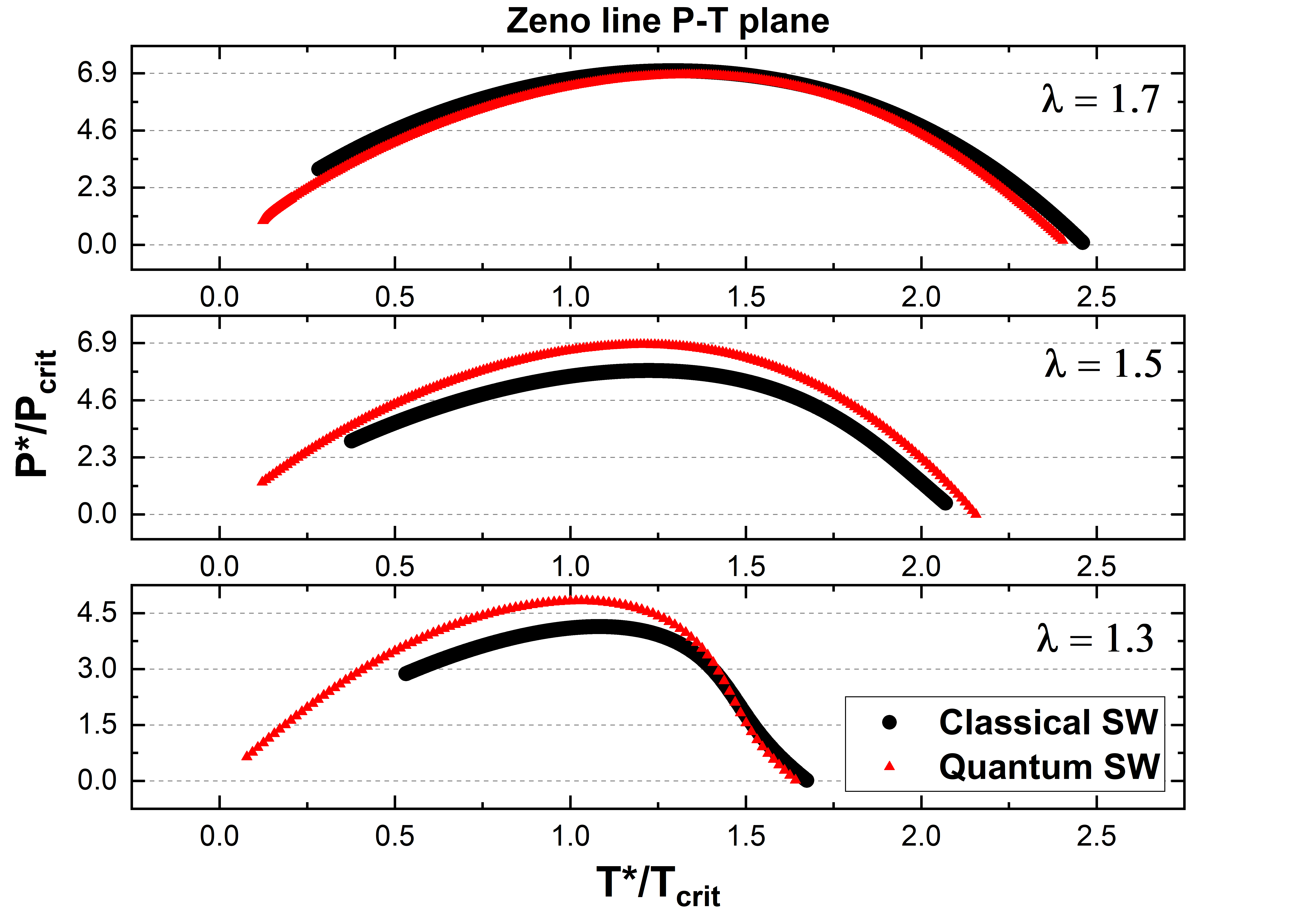}\\
\hrulefill \\
\caption{Comparison of the Classical and Quantum Zeno lines for $\lambda^* = 1.3$ (bottom), $\lambda^* = 1.5$ (middle), and $\lambda^* = 1.7$ (upper).}
\label{Zenoline}
\end{figure}

The line often associated in the literature with TG and other geometric formulations of thermodynamics,  with the ideal gas behavior \cite{Quevedo2015,Quevedo2022}, was also analyzed. This line is constructed with  the set of thermodynamic states with zero curvature, $R^*(T^* , \rho^*)=0$. For all interaction ranges considered, each of the states that satisfies such condition are located  in a very high density region, beyond the domain of validity of the equations of state considered. For this reason, these results are not shown in the present work. Nevertheless, the fact that the same qualitative behavior was observed for all ranges and both classical and quantum fluids suggests that the  line $R=0$, lies in the high–density regime and, therefore, its interpretation as an ideal line in the sense of following an ideal gas-like behavior should be carefully revised and explored.

In the absence of ideal lines emerging from the thermodynamic geometry framework, we explored another well-studied alternative, the Zeno line, defined as the set of thermodynamic states for which the compressibility factor equals one, $Z \equiv P/\rho k_B T$, corresponding to an ideal gas, to enable a comparison, at least qualitatively, of its behavior with that of the ideal line $R=0$. The result is presented in Fig. \ref{Zenoline}. The behavior observed for these Zeno  lines differs substantially  from that of the $R=0$ lines, and instead resembles the behavior of the Widom lines. In particular, they exhibit a non-monotonic trend: increasing at low temperatures, reaching an extremum, and decreasing at higher temperatures.  As in the case of the Widom lines, the discrepancies between classical and quantum results diminish as the interaction range increases and at low temperatures.

\begin{figure}[h]
\centering
\includegraphics[width=\columnwidth]{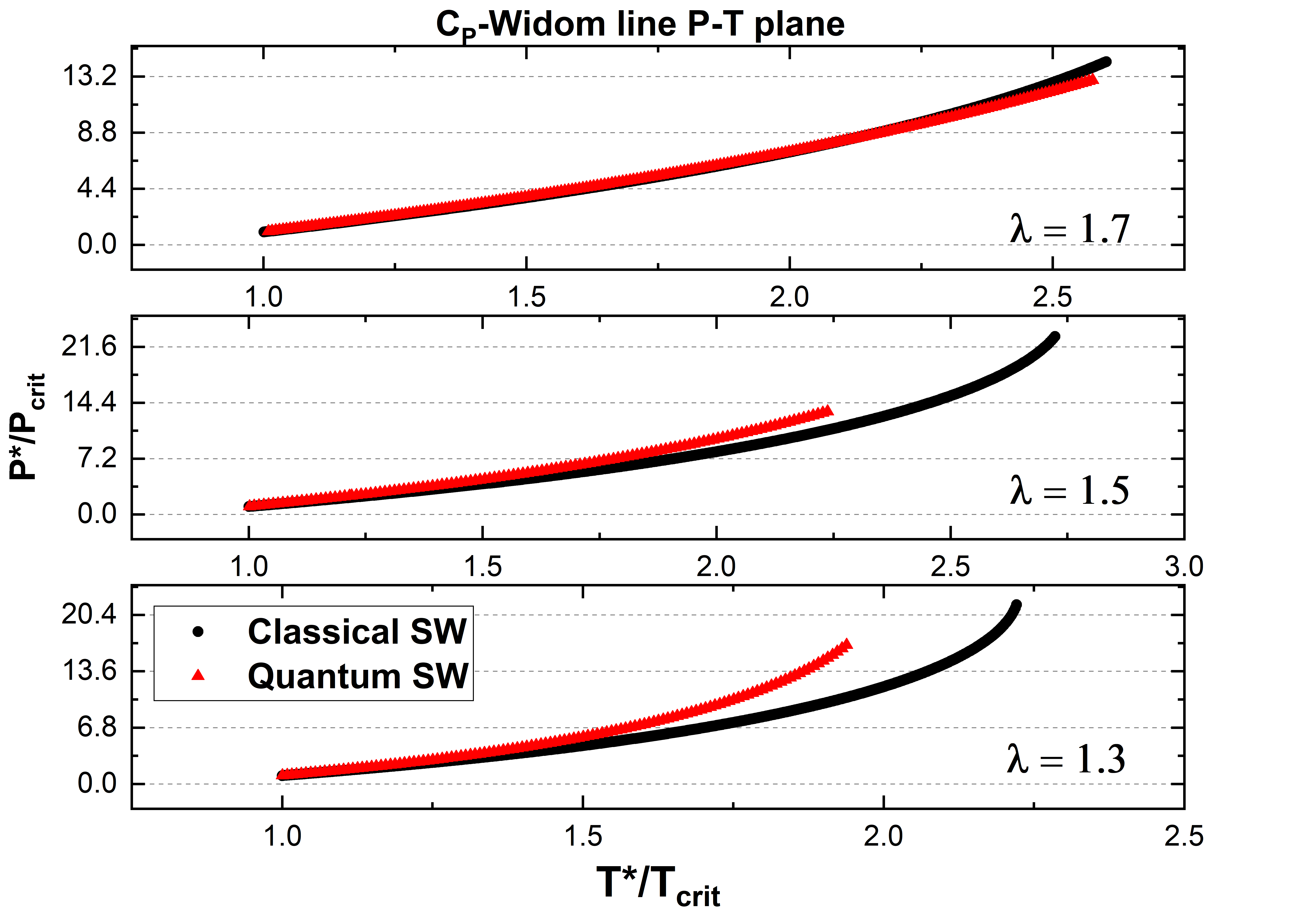}\\
\hrulefill \\
\caption{Comparison of  the SW Classical and Quantum $C_P$-Widom lines for $\lambda^* = 1.3$ (bottom), $\lambda^* = 1.5$ (middle), and $\lambda^* = 1.7$ (upper).}
\label{WidomCp}
\end{figure}

As illustrated in Figure \ref{WidomCp}, the classical and quantum Widom lines, defined by the loci of the maxima of $C_p$, exhibit distinct behaviors in the different potential ranges considered. Discrepancies are more pronounced for shorter-range interactions; consequently, at a given temperature, the quantum crossover occurs at a pressure higher than that predicted by the classical model, with the difference becoming particularly significant at the shortest ranges.

\begin{figure}[h]
\centering
\includegraphics[width=\columnwidth]{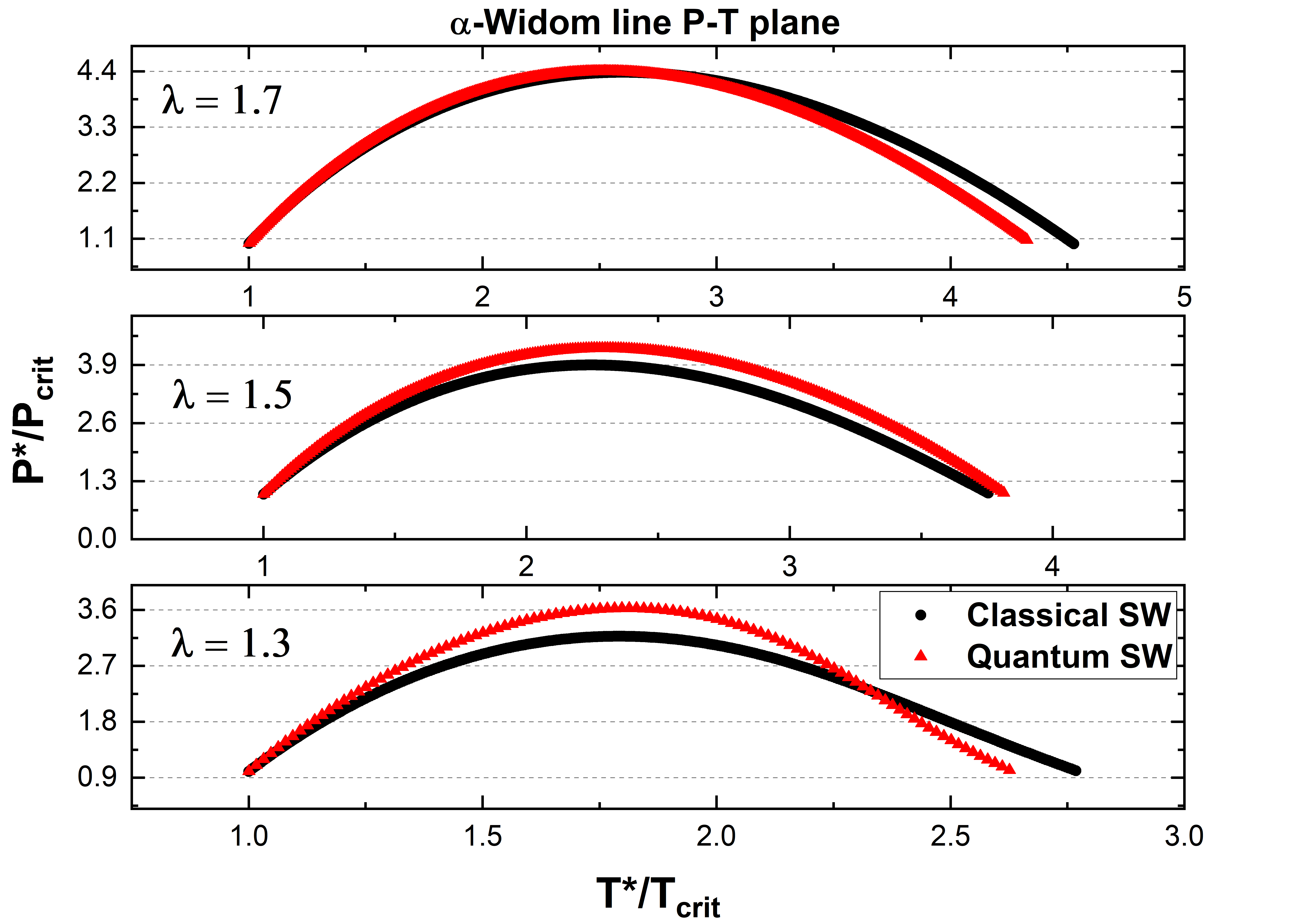}\\
\hrulefill \\
\caption{Comparison of the Classical and Quantum $\alpha$-Widom lines for $\lambda^* = 1.3$ (bottom), $\lambda^* = 1.5$ (middle), and $\lambda^* = 1.7$ (upper).}
\label{Widomalfa}
\end{figure}

Similarly, Figure \ref{Widomalfa} presents the $\alpha$-Widom lines. As expected, these lines span a region broader than those associated with other response functions or with scalar curvature $R$. As in previous cases,  discrepancies are most pronounced for short-range interactions, with a clear tendency for the classical and quantum results to converge as the range increases.

\begin{figure}[h]
\centering
\includegraphics[width=\columnwidth]{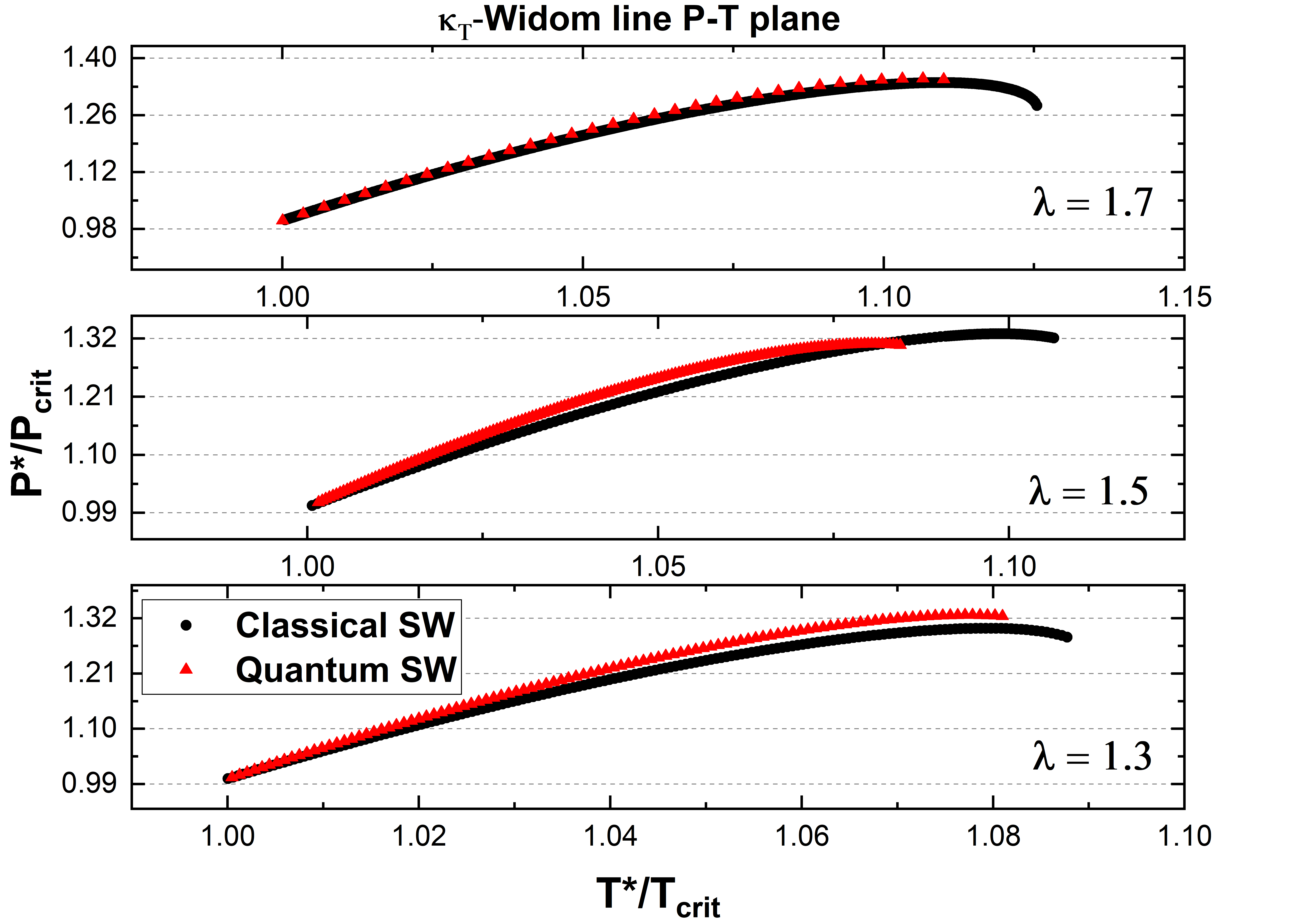}\\
\hrulefill \\
\caption{Comparison of  the Classical and Quantum $\kappa_T$-Widom lines for $\lambda^* = 1.3$ (bottom), $\lambda^* = 1.5$ (middle), and $\lambda^* = 1.7$ (upper).}
\label{Widomkappa}
\end{figure}

Interestingly, for the case of the Widom lines associated with the isothermal compressibility, presented in Figure \ref{Widomkappa}, although some differences between the classical and quantum results are evident   across the considered ranges,  the overall behavior is largely insensitive to both regimes. This suggests that, when accounting for the quantum nature of the fluid, changes in density (related to the compressibility) are less sensitive than changes in temperature (related to the heat capacity and thermal expansion coefficient).


	
\section{Conclusions}

In this work, we have investigated the impact of quantum contributions on thermodynamic geometry and Widom-line behavior in supercritical fluids within a perturbative framework. By comparing classical and quantum descriptions across multiple  interaction ranges, we have assessed how quantum effects modify both geometric and response-function-based indicators beyond the critical point.

Our analysis of the scalar curvature isotherms shows that quantum effects smooth the anomalies typically observed near the critical point  and induce a shift of the extrema toward lower densities for short-range potentials. Despite these quantitative changes, the critical behavior itself remains unchanged: the critical exponents are consistent with mean-field predictions for all fluids considered, as expected for a perturbative theory.

The behavior of the $R$-Widom lines reveals persistent differences between the classical and quantum descriptions across the entire range of interaction ranges examined. Due to the higher-order derivative structure of the scalar curvature $R$, these discrepancies are plausibly attributable to the intrinsic sensitivity of $R$ to quantum corrections. Unlike other Widom lines, no clear trend toward convergence between classical and quantum results is observed within the explored ranges, suggesting that any such convergence, if present, may only occur for substantially longer interaction ranges, non accessible by the considered quantum SW equation of state.

Thermodynamic lines defined by the condition $R=0$, often associated with ideal-gas-like behavior, were also examined. For every range considered, these lines were located in a high-density region, outside the validity domain of the equation of state employed. Consequently, a quantitative assessment is not feasible in the current scheme; nevertheless,  the consistency of this behavior across all ranges suggests that the interpretation of the $R=0$ line as a ideal-gas-like behavior should be reconsidered in this context. 

For qualitative comparison,  the Zeno line, defined by all thermodynamic states that meet the condition $Z=1$, was analyzed. Its behavior differs markedly from that of the $R=0$ line, resembling instead that of Widom lines, exhibiting a non-monotonic temperature dependence. As observed for Widom lines, the differences between classical and quantum predictions along the Zeno line decrease with increasing interaction range and at low temperatures.

A systematic comparison of Widom lines associated with different thermodynamic response functions (and scalar curvature ) highlights a strong dependence on both the interaction range and the specific thermodynamic quantity considered. Widom lines derived from $C_P$ and from the thermal expansion coefficient $\alpha$ exhibit significant classical–quantum discrepancies for short-range potentials, with a clear tendency towards convergence as the interaction range increases. This feature is in agreement with previous results for the QSW system \cite{Trejos2012,serna2016molecular} where quantum effects arising in the HS repulsive term are very relevant, also in the behavior of the perturbation terms, since $a_{1_\text{PI}}$ significantly differs from the classic value for short-range values of $\lambda^*$. In contrast, Widom lines associated with the isothermal compressibility $\kappa_T$, show a relatively minor sensitivity to quantum effects, suggesting that density fluctuations are less affected by quantum corrections than temperature-related fluctuations within the present framework.

Overall, these results suggest  that quantum effects can significantly  modify thermodynamic features, particularly those associated with thermodynamic geometry, while  the universal critical behavior remains unchanged at the current  level of approximation. The interplay between interaction range, quantum corrections, and specific thermodynamic observables appears to play a key role in determining the full extent of classical–quantum deviations in bulk systems, motivating further investigation beyond the perturbative regime considered here. The other effect to be considered is the behavior under confinement, where strong effects have been reported previously in connection also with the effective dimension of the system \cite{Trejos2013, Contreras2021, Gil-Villegas2025}.

\acknowledgments{J. L. López-Picón was partially supported by Secretaría de Ciencia, Humanidades, Tecnología e Innovación (SECIHTI) and University of Guanajuato (UG).\\
L. F. Escamilla-Herrera acknowledge support from SECIHTI through postdoctoral grant: Estancias Posdoctorales por México para la Formación y Consolidación de las y los Investigadores por México (CVU: 230753) and UG through grant 401/2025 of Convocatoria Institucional de Investigaci\'on Cient\'ifica 2025.\\ 
A. Gil-Villegas acknowledge support from SECIHTI (Grant CBF2023-2024-3350)\\
J.~Torres-Arenas gratefully acknowledges the Departamento de F\'isica Aplicada of the Universidade de Vigo for the facilities and support provided during his sabbatical stay, during which, part of this work was carried out. He also acknowledges the financial support from the Universidad de Guanajuato and SECIHTI (through SNII), which made this stay possible.}

\bibliographystyle{apsrev}
\bibliography{ThermoGeo}

\end{document}